\documentclass{aastex}
\usepackage{spr-astr-addons}
\usepackage{url}

\RequirePackage{color}

\begin{document}

\title{Sunspot Drawings at Kodaikanal Observatory: A Representative Results on Hemispheric Sunspot Numbers and Area Measurements}

\shorttitle{Sunspot Drawings at Kodaikanal Observatory}
\shortauthors{Ravindra et~al.}

\author{Ravindra B.\altaffilmark{1}} \author{Kumaravel Pichamani\altaffilmark{1}} \author{Selvendran R.\altaffilmark{1}} 
\author{Joyce Samuel\altaffilmark{2}} \author{Praveen Kumar\altaffilmark{3}}, \author{Nancy Jassoria\altaffilmark{4}} \and \author{Navneeth~R.~S\altaffilmark{5}}
\email{ravindra@iiap.res.in}

\altaffiltext{1}{Indian Institute of Astrophysics, Koramangala, Bengaluru - 560 034, INDIA, {\bf email:} {\em ravindra@iiap.res.in}}
\altaffiltext{2}{The American College, Tallakkulam, Madurai - 625002, India}
\altaffiltext{3}{Aix-Marseille University, 6 Traverse Notre Dame Des Grâces, Marseille-13014, France}
\altaffiltext{4}{Punjabi University, Patiala - 147002, Punjab, INDIA}
\altaffiltext{5}{Amrita Vishwa Vidyapeetham, Amrita School of Arts and Sciences, Amritapuri Campus,  Kollam - 690525, INDIA}

\begin{abstract}
The importance of the periodicity in sunspot appearance was well recognized by the mid of 19th century. Several observatories around the globe have made the record of sunspots in the form of drawings and preserved them safely for posterity. At the Kodaikanal Observatory (KO), the sunspot observations have begun in 1905. In those times observations were recorded on photographic plates and after the development of those plates in the laboratory, the drawings of the same were made on the Stonyhurst grids. In these drawings, called {\em sun charts}, different features on the sun's disk, e.g., sunspots, plages, filaments, prominences, etc. were clearly identified and visually marked with different colors.  We have collected 111 years of sunspot drawing spanning over 10 solar cycles. These sunspot drawings were carefully stitched to make bound volumes, each for every 6-months. The drawings are kept at the Kodaikanal library for scientific use and analysis. In this article, we describe briefly the process of drawing, methods of counting sunspot numbers and measurement of sunspot area using square grids. We have collected the data for the Northern and Southern hemispheres separately. From the collected data, we compute the sunspot number and area and compare it with Royal Greenwich Observatory (RGO) and Sunspot Index and Long-term Solar Observations (SILSO) data. The results show that the measurement of the sunspot number is underestimated by about 40\%.  The KO monthly averaged sunspot number data of both hemispheres is normalized with the RGO monthly averaged total sunspot number data. This procedure provided a good correlated data set that can be used for further scientific work in the future. The sunspot area data from KO is slightly underestimated in the first 5-cycles and seems mildly overestimated in the later 5-cycles. These data sets are valuable additions to the existing data of sunspot records.
\end{abstract}


\setcounter{page}{1}

\section{Introduction}
The most prominent observable features on the sun are sunspots and sunspot groups. The sunspots have the longest record in history \citep{2010LRSP....7....1H,2015LRSP...12....4H}, for investigating subtle and interesting long-term trends in solar activity cycle. More specifically, long historical records are useful to understand the physics behind the solar activity and the solar dynamo believed to be operating in the solar interior to generate large scale solar magnetic fields which in turn are responsible for sunspots appearing on the solar surface \citep{2010LRSP....7....3C}.  To keep a record of solar activity, the sunspot number \citep{1961says.book.....W} and group numbers \citep{1998SoPh..179..189H, 2017LRSP...14....3U} are mainly used. Its record goes back to the early telescope era, the beginning of the 17th century. Those records are based on full-disc solar drawings and sketches.

The systematic record-keeping of sunspots on the charts has begun during the Galileo Galilee time by Harriot, Johann Fabricius (1611) and many others. Although, the nature and cause of various features observed on the surface of the sun were barely clear to anyone for a long time, the curiosity-driven systematic observations and book-keeping continued. In those times the drawing would be made by projecting the solar image on the wall or on the sheet of the paper, directly. The activity of making solar drawings was continued in the 19th century by William Herschel, Richard Christofer Carrington and others \citep{1863london.book.....C}. Many of the long-term trends in the sunspot behavior were inferred from these charts and Carrington maps. However, by the end 
of 19th century the availability of the photographic plates/films made the recording process convenient and easy \citep{2013A&A...550A..19R}. 

Johann Caspar Staudacher, an amateur astronomer made solar drawings starting from 1749 -- 1796 \citep{2008SoPh..247..399A}.  Recently from these drawings, the locations of sunspots, sunspot numbers, etc. were extracted \citep{2009ApJ...700L.154U}. His drawings are preserved in the Astrophysikalisches Institut Potsdam library. Samuel Heinrich Schwabe has tracked the sunspots for more than 40-years \citep{2011AN....332..805A}. Richard Christopher Carrington also made sunspot observations starting from 1853. The Friedrich Wilhelm Gustav Sp\"{o}rer started the solar observations in 1860 \citep{1861AN.....55..289S} and continued till 1894. He collected about 33-years of sunspot drawings. From Rundetarn in Copenhagen between 1761 to 1776, Christian Horrebow made regular observations of sunspots  \citep{2019SoPh..294...77J}. The drawings from six observatories in China are digitized recently \citep{2019SoPh..294...79L} and the data is available from 1915 up to 2015. In Japan, the white-light observations started in 1918. From 1938 -- 1998 the sunspots and faculae were sketched on the paper by projecting the image on it.  After 1990, they have imaged the sun in white-light using electronic devices \citep{2016ASPC..504..313H}. At the Mt. Wilson Observatory, the white-light and Ca-K observations of the sun started during the beginning of the 20th century \citep{1985SoPh..100..171H}. These observations were made with the 150-foot tower telescope and photographic plates.  Full-disk drawings of the sunspots were made at Boulder, Sacramento Peak \citep{1981SoPh...69..411C} and Holloman observatories starting from 1948 till 1992 and combined them to make one set of data.   

After 1870 many observatories started the recording of the sun's image and sunspots in photographic plates. Greenwich observatory started its photographic observations of the sun in 1874 \citep{2013SoPh..288..117W}.  They have collected the white-light 
image of the sun taken in photographic plates starting from April 1874 until December 1976. They have also compiled data obtained from several observatories around the world such as Greenwich, Herstmonceux, Cape of Good Hope, Dehra Dun, Kodaikanal, and Mauritius. Later, Debrecen Heliophysical Observatory collected about 150,000 photoheliograms since 1958 \citep{2016SoPh..291.3081B}. 

With all the available sunspot drawings, Wolf \citep{wolf1851} and Wolfer \citep{1902PA.....10..449W} derived relative sunspot numbers to confirm the periodicity in the number of sunspots seen on the sun. Carrington has made maps of the full sun by considering one image per day and put them in latitude and longitude grid \citep{carrington1863}. These maps showed the full 360$^{\circ}$ view of the sun and also provided the activity level on the sun. 

Apart from sunspot numbers, sunspot area also measured from the sunspot data and which clearly showed the solar activity with an average period of 11-years \citep{2010LRSP....7....1H, 2015LRSP...12....4H}.
The discovery of magnetic fields in sunspots, the opposite polarity of leader and follower spots, polarity reversal over 11-years, cyclic behavior of polar fields, butterfly diagram of solar activity cycle, periodicity in sunspot number, etc. prompted Babcock \citep{1961ApJ...133..572B} to put forth the hypothesis of the solar dynamo.

The sunspot area measurements are made for both the hemisphere separately to study the asymmetry parameter \citep{2009SoPh..254..145L}, to identify 
the location of the sunspots to make butterfly diagram \citep{2009SoPh..255..143A} and many more. Though the sunspot number count had started before everything else, no separate measurements were carried out for the hemisphere sunspot number count. (see also: \cite{2002A&A...390..707T}). 

Kodaikanal has a history of making solar observations in different wavelengths including in white-light since the year 1905. The drawings of these images were also made on the paper, with information about sunspots, plages and filaments being distinctly recorded. In this paper, we describe the sunspot drawings available at Kodaikanal Observatory and how these drawings are used to extract the sunspot number and area separately for both the hemisphere.  The results obtained from these drawings are compared with those from Royal Greenwich Observatory (RGO), Kanzoelhoe Solar Observatory (KSO) and Sunspot Index and Long-term Solar Observations (SILSO) sunspot number and area data. Further, for a homologous comparison, we normalized the Kodaikanal hemispheric sunspot number data and rescaled it to the RGO sunspot number.  This provided an important hemispheric sunspot number data set for further study. We end the paper with a discussion of the results and conclusions.

\section{Observations and Drawings}
At the Kodaikanal Observatory (KO) the white-light images of the sun have been recorded in photographic plates since 1905. The image of the sun is projected on the screen with the help of a 15 cm lens placed on the front-side of the equatorial telescope. The projected image of the sun (10.2 cm in radius) is recorded on the 25.4 $\times$ 25.4 cm$^{2}$ photographic plate. More details about the telescope, observations, and preservation of the photographic plates/films can be found in  \cite{2013A&A...550A..19R}. A few years ago the digitization of these photographic plates is taken up and finished. Details of the digitization, data calibration and the data availability for the scientific use are described in \cite{2013A&A...550A..19R} and \cite{2017A&A...601A.106M}. 

\begin{figure*}[!h]
\begin{center}
\includegraphics[width=0.4\textwidth]{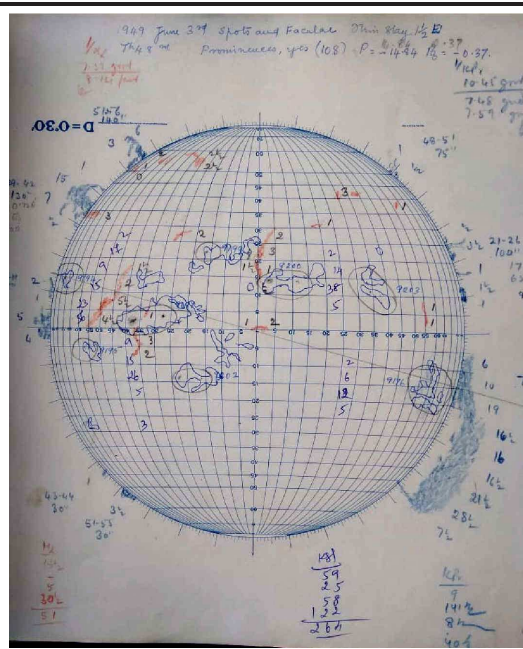}
\end{center}
\caption{The Stonyhurst map of the sun over which sunspots (drawn in pencil), plages (drawn with blue-colored pencil), filaments (drawn with red-colored pencil), and prominence (drawn with blue colored pencil) are shown. The KKL number is assigned for each sunspot group as shown in the map. The total sunspot number, area, and much more information are also shown in the chart.}
\label{fig:1}
\end{figure*}

In addition to the photographic plates, at the KO the drawings of sunspots, plages and filaments were also made on the Stonyhurst latitude and longitude grid (Figure~\ref{fig:1}). This was done by projecting the white-light sunspot image on to the Stonyhurst latitude and longitude grid. The grid size is 5$^{\circ}$ in both the latitude and longitude direction. These drawings are made with sharp-edged HB pencils. On the same grid, the Ca~K images recorded in the photographic plate is projected using the lens and light arrangement. The positions of the photographic plate and the grid are fixed. The lens is moved back and forth to focus and to make the size of the projected image the same as the grid image. Then using the blue colored ball pen, the borders of the plages are sketched.  Again, the image obtained in H$_{\alpha}$ is projected on to the same grid using the same lens and lamp arrangement.  The solar filaments image are clearly seen in the projected image. The edges of the filament feature are drawn using a red colored pencil.  The limb prominence images are obtained in Ca~K wavelength using the disk which blocks the photosphere light entering into the spectrograph.  The limb prominences are drawn with blue colored pencils. 

The central longitude (L$_{0}$) and latitude (B$_{0}$) values change with the time. 
The charts are made for each 0.5$^{\circ}$ B$_{0}$ values. The mean B$_{0}$ line is drawn for that month using the pencil in the East-West direction passing through that value of B$_{0}$ on the grid. In order to align the grid with the image of the day, the East-West line of the sky which is present in the image aligned with the B$_{0}$ line in the grid.  Once it is aligned, all the sunspots are drawn with HH marker pencil till the year 1995 and after that with HB marker pencil is used till today. The observer at KO makes these drawings/sketches of the features on the Stonyhurst grid. Generally, the same observer makes these drawings every day. Only during the absence of that particular observer, other skilled observers will take up this work until the main observer resume the work.  If there are two or more photographic plates/films available then the observer chooses the better plate/film for drawing. These drawings are made into booklets called ``sun charts'' and kept in the KO library for scientific use. Each booklet has six months of data and for one solar cycle, there are 22 such books available. 

These {\it sun charts} are used to count the sunspot numbers and measure sunspot areas. Many of the numbers written on it shown in the corresponding color. This is to show each color corresponds to the numbers associated with the same colored features. For the sunspot number counts, we used daily Wolf sunspot number obtained by an individual observer from the same station images and is defined as, 
\begin{equation}
R_{n} = k.(10.g + n)
\end{equation}

where, {\it k} is the correction factor for the different observatory, {\it g} is the number of sunspot groups observed on the solar disk and {\it n} is the total number of individual sunspots observed on the disk. For the present counting of sunspot number, we have taken {\it k} as 1. We have separated the Northern and Southern hemispheres and recorded the sunspot numbers.  The sunspot number is recorded starting from 1905 till 2016. The exercise is repeated randomly by a third person to verify the consistency in recorded numbers.  We cross-checked the whole month's data if there is a discrepancy in the recorded number. In a similar way, the observer recorded the total sunspot number on the top of the chart every day. We rechecked our numbers with the entered one to ensure the accuracy against the procedural lapses. 

We have also measured the sunspot group area using two-millimeter square grid arranged in a circular pattern. Each grid is of the size 2~mm. We kept 
the grid on the {\it sun chart} by matching the limb of the square grid to that of {\it sun chart}. Once it is matched, then using the grid we have counted the number of square grids the sunspot group occupied.  Even, we have counted the half and 3/4$^{th}$ boxes as well. There is a conversion table available for the number of boxes the sunspot group occupied which can be converted to a millionth of a hemisphere (mh). This also uses the angle made by the centroid of the group of sunspots with the equator. This angle can be used to correct the projection effect. We have measured the heliocentric angle made by the sunspot group centroid with respect to the center of the solar disk. Then using the table we converted the number of square grid to the millionth of hemisphere \citep{2002JBAA..112..353M}. For example, if the sunspot group has covered 10 boxes and it is located at a heliocentric angle of 30 degrees then the area covered by the group is 722 millionth of a hemisphere. 

The KO has obtained the white-light observations of sunspots from 1905 till today in photographic plates and films. These images are further recorded on the Stonyhurst grid using a pencil. Figure~\ref{fig:1} depicts the drawing of sunspots and other features made on the Stonyhurst grid. Each sunspot group is marked with a box, the KKL number is written in black pen on the top of the box. All drawings are made into book format and made it is available to the scientific users. We have used those {\it sun charts} to get the Wolf number and area occupied by the sunspot group separately for the Northern and Southern hemispheres.

\section{Results}
The sunspots are drawn on the Stonyhurst grid whose size changes with the season. Even the visible pores are drawn on the charts. The sunspot groups are counted separately for both the hemispheres. The Wolf's sunspot number is recorded in the Excel sheet along with the date and time of the image obtained from the photographic plates. Along with this, the projection corrected sunspot group area is also recorded separately for both the hemispheres. In the following, we will present the results of the recorded data sets. 

\subsection{Wolf's Sunspot Number}

\begin{figure}[!h]
\begin{center}
\includegraphics[width=0.45\textwidth]{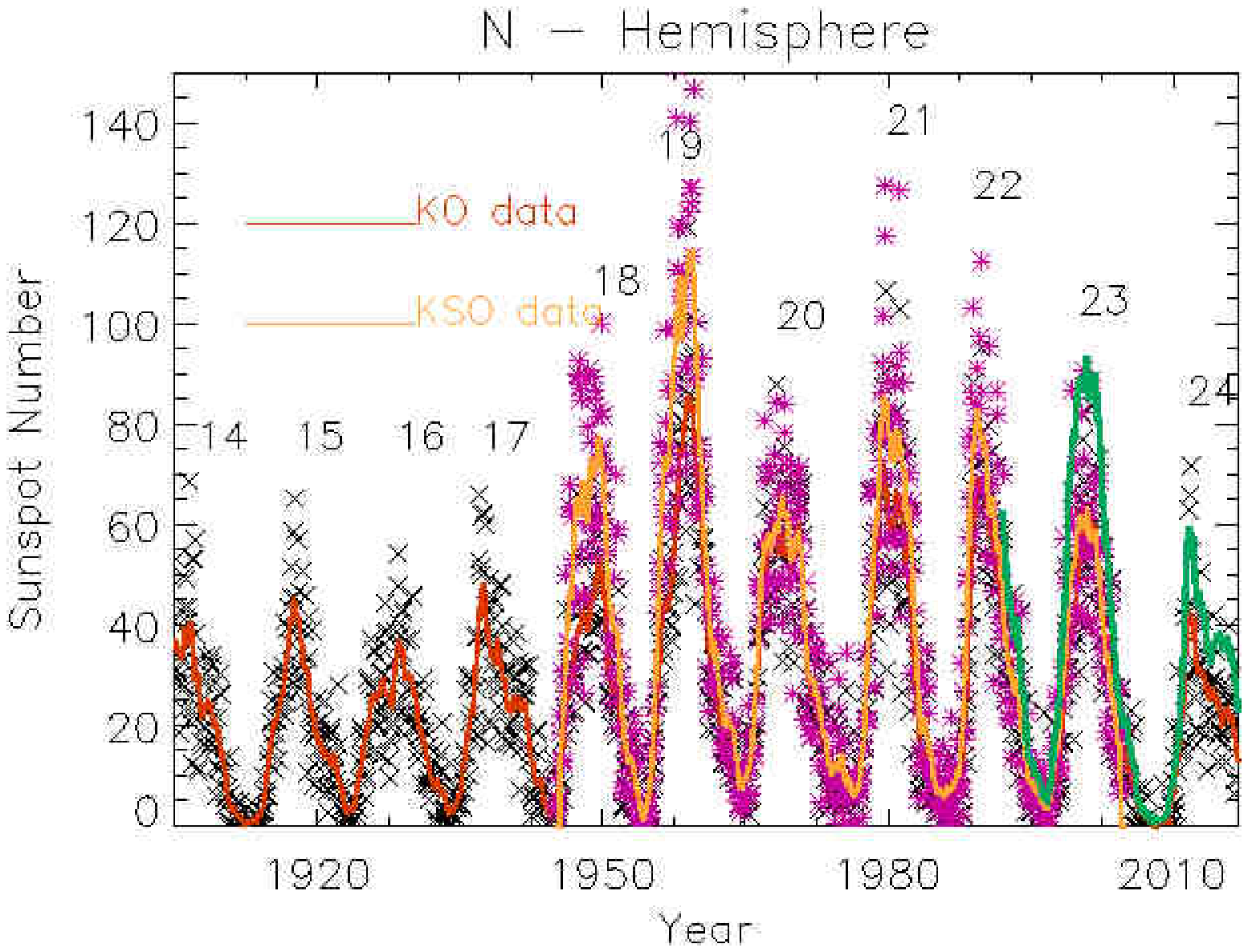} \\
\includegraphics[width=0.45\textwidth]{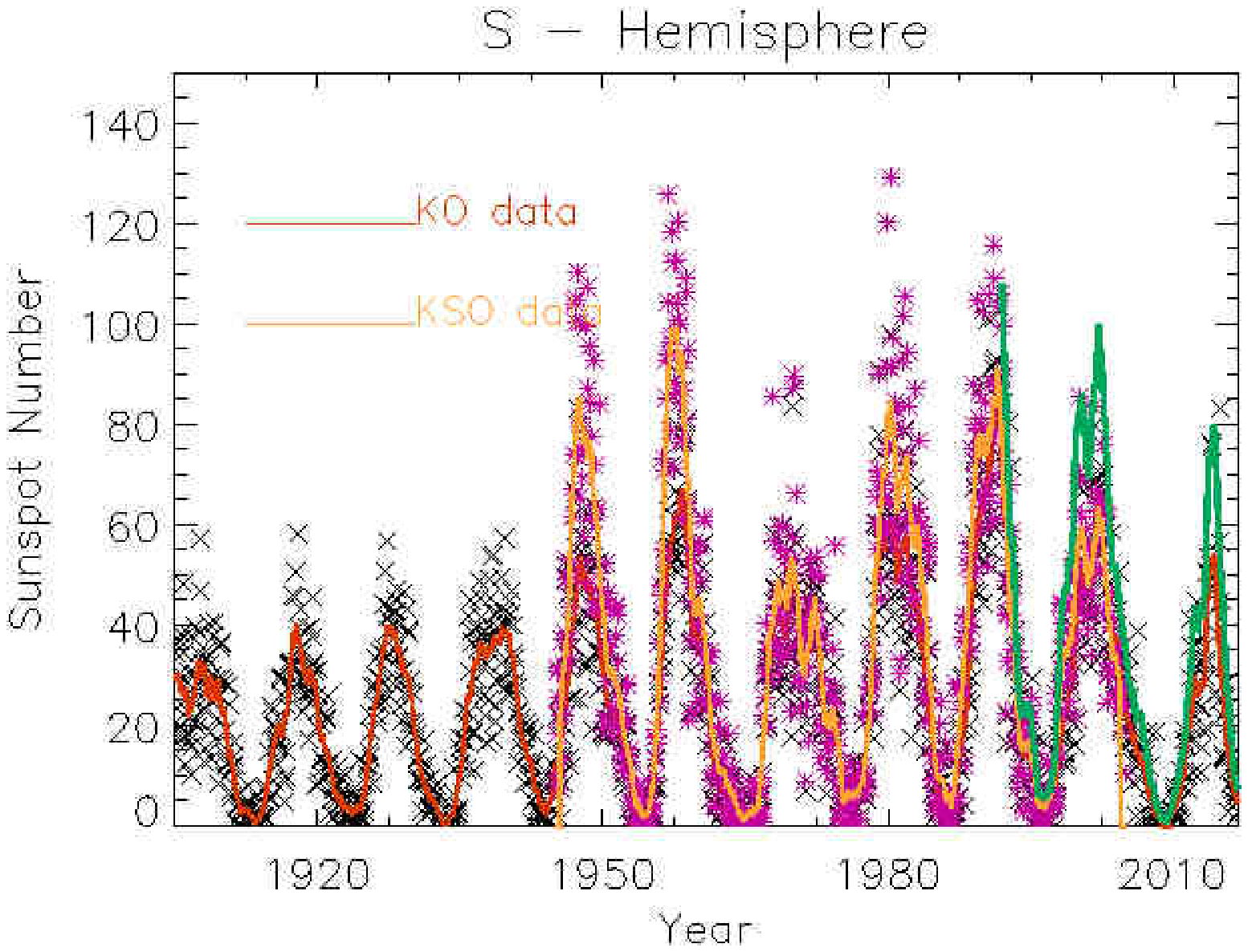} \\
\end{center}
\caption{Top: The monthly averaged Wolf's sunspot number is plotted against the time of observations for the Northern hemisphere of the sun. The solar cycle starts from number 14 and covered up to the ongoing cycle 24. In the plot, the black cross ($\times$) symbol represents the monthly averaged Kodaikanal observatory data and the pink-colored star ($\ast$) symbol represents the Kanzoelhoe Solar Observatory sunspot number data. The red-colored solid curve and the orange-colored curve represent the 13-months smoothed KO and KSO data respectively. The green-colored curve represents the 13-months smoothed SILSO data. Bottom: Same as top-side plot but for the Southern hemisphere of the sun.}
\label{fig:2}
\end{figure}

The relative sunspot number defined by the R. Wolf (1856), uses the total number of sunspots and groups seen 
on the visible hemisphere of the sun. For the present study, we fix the value of scale factor to one.  At the 
KO we followed the same procedure of Wolf to compute the sunspot number, except that the data is only from the 
KO and the number of observers who cross-examined the groups and sunspots are a few. In addition to this, the 
sunspot number for both hemispheres is computed separately.  We averaged the sunspot number data for each 
month and for both the hemispheres separately. Figure~\ref{fig:2} shows the monthly averaged sunspot number 
for Northern (top) and Southern (bottom) hemisphere plotted against the observations year. The plot did not
 cover cycle 14 and 24 completely. To compare the KO sunspot number data we have overplotted the sunspot number
data obtained from combined observations from Konzelhohe solar observatory (KSO), Austria  and the 
Skalnat{\'e} Pleso Observatory, Slovak Republic published by \cite{2006A&A...447..735T} . They have 
covered the data from solar cycle 18 -- 23. They have separately counted the sunspot number for the Northern 
and Southern hemisphere and normalized their values to the total sunspot number published by Greenwich 
Observatory.  The red, orange and green colored lines are the 13~months smoothed data for the KO, KSO and 
SILSO data set. Even the SILSO (green color) is normalized to the RGO total sunspot number data. Comparing 
the KO data with the SILSO and KSO data, we find that the KO data values are systematically lower than 
the two. This is clearly seen in the monthly averaged data as well as in 13~months smoothed data sets.  
However, the pattern around the peak appears to be the same in the KO data with other data sets. A similar pattern 
is seen in the southern hemisphere data of KO and it is systematically lower than the other two data sets. 

\begin{figure*}[!h]
\begin{center}
\includegraphics[width=0.5\textwidth]{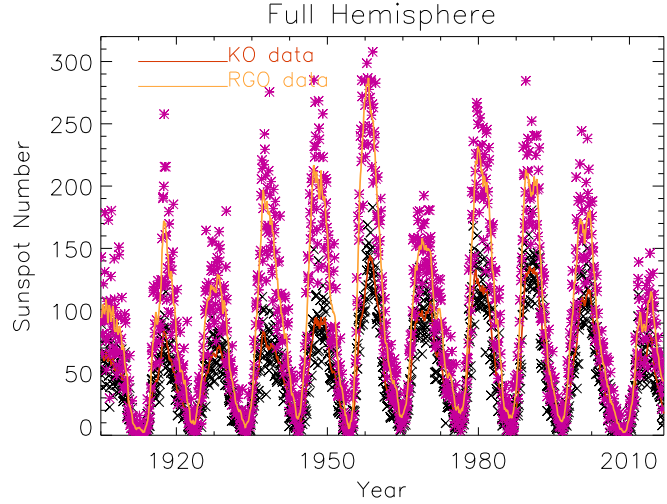} \\
\end{center}
\caption{The total monthly averaged sunspot number is plotted for cycle 14 through 24. The black cross (x) represents the Kodaikanal sunspot number data and the pink star ($\ast$) represents the Royal Greenwich data. The red and yellow-colored solid line represents the 13-months smoothed KO and RGO sunspot data respectively.}
\label{fig:3}
\end{figure*}

Figure~\ref{fig:3} shows the plot of the total monthly averaged sunspot number extracted from KO and overplotted with the RGO monthly averaged data set. The red and orange-colored curves represent the 13-months smoothed KO and RGO data sets. Once again the plot shows the KO sunspot number data is systematically lower than the RGO data sets.  However, the pattern appearing at the peak of the cycle is similar in almost all the cycles. 

\begin{figure*}[!h]
\begin{center}
\includegraphics[width=0.5\textwidth]{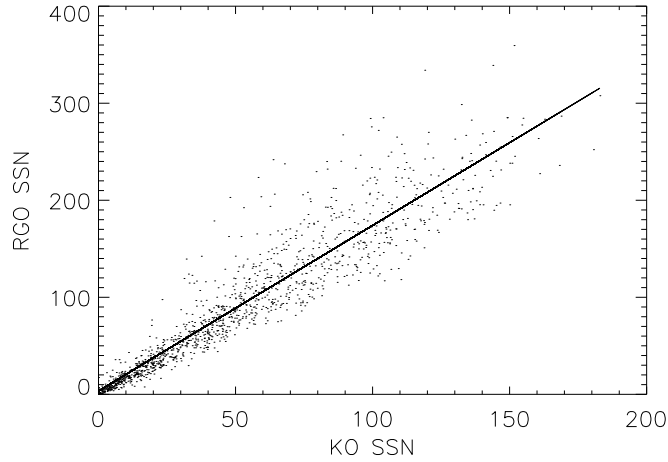} \\
\end{center}
\caption{A plot of monthly averaged total sunspot number obtained from Greenwich data is plotted against the Kodaikanal data. The solid line is a linear least-square fit obtained to the data points.}
\label{fig:4}
\end{figure*}

A first look at the sunspot number profile seen in Figure~\ref{fig:3} approximately indicates that the amplitude of each cycle measured from KO data is about 2 times smaller than the RGO data. The systematic underestimation of sunspot number measured from KO data is depicted in the scatter plot between RGO and KO total sunspot number (Figure~\ref{fig:4}). The scatter is small at small sunspot number and it is large as the sunspot number value increases. The linear 
least-square fit to the data points provide a following relationship between the two data sets.
\begin{equation}
RGO\_SSN =  1.7(\pm 0.017) \times KO\_SSN +  3.46 (\pm 1.107)
\end{equation}
The relationship shows that the KO is 1.7 times smaller than the RGO data set. Just multiplying the coefficient 1.7 to the KO sunspot number data will lead to the underestimation or the overestimation of the values compared to the RGO number. Hence, it may be better to normalize the hemispheric sunspot number data with the whole disk sunspot number and then multiply that coefficient with the RGO data to get a comparable sunspot number with other data sets as has been described in \cite{2002A&A...390..707T, 2006A&A...447..735T}.  

\subsection{Normalized Sunspot Numbers}
Though the profile of sunspot number as a function of time is similar to the RGO measured sunspot number, but the amplitude is highly underestimated. To make 
it comparable to the RGO sunspot numbers we followed the method of \cite{2002A&A...390..707T}. In short, we first normalized the Northern and Southern sunspot numbers to the full-disk sunspot number and obtained the coefficient `Nc' and `Sc'. These coefficients are multiplied with the RGO full-disk sunspot measurements. For this purpose, we used the monthly averaged sunspot number data of both the KO and RGO. 

\begin{figure*}[!h]
\begin{center}
\includegraphics[width=0.45\textwidth]{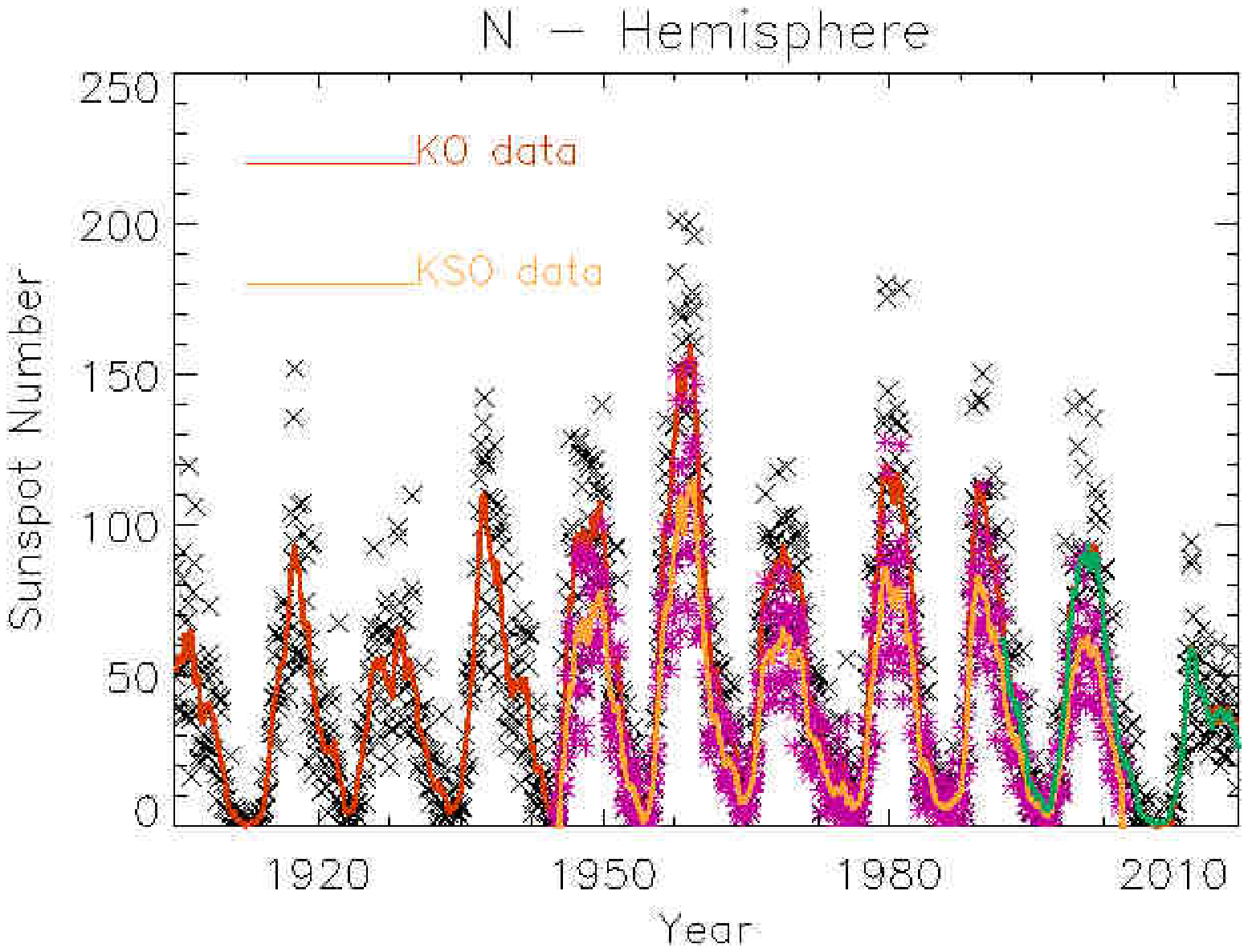}\includegraphics[width=0.45\textwidth]{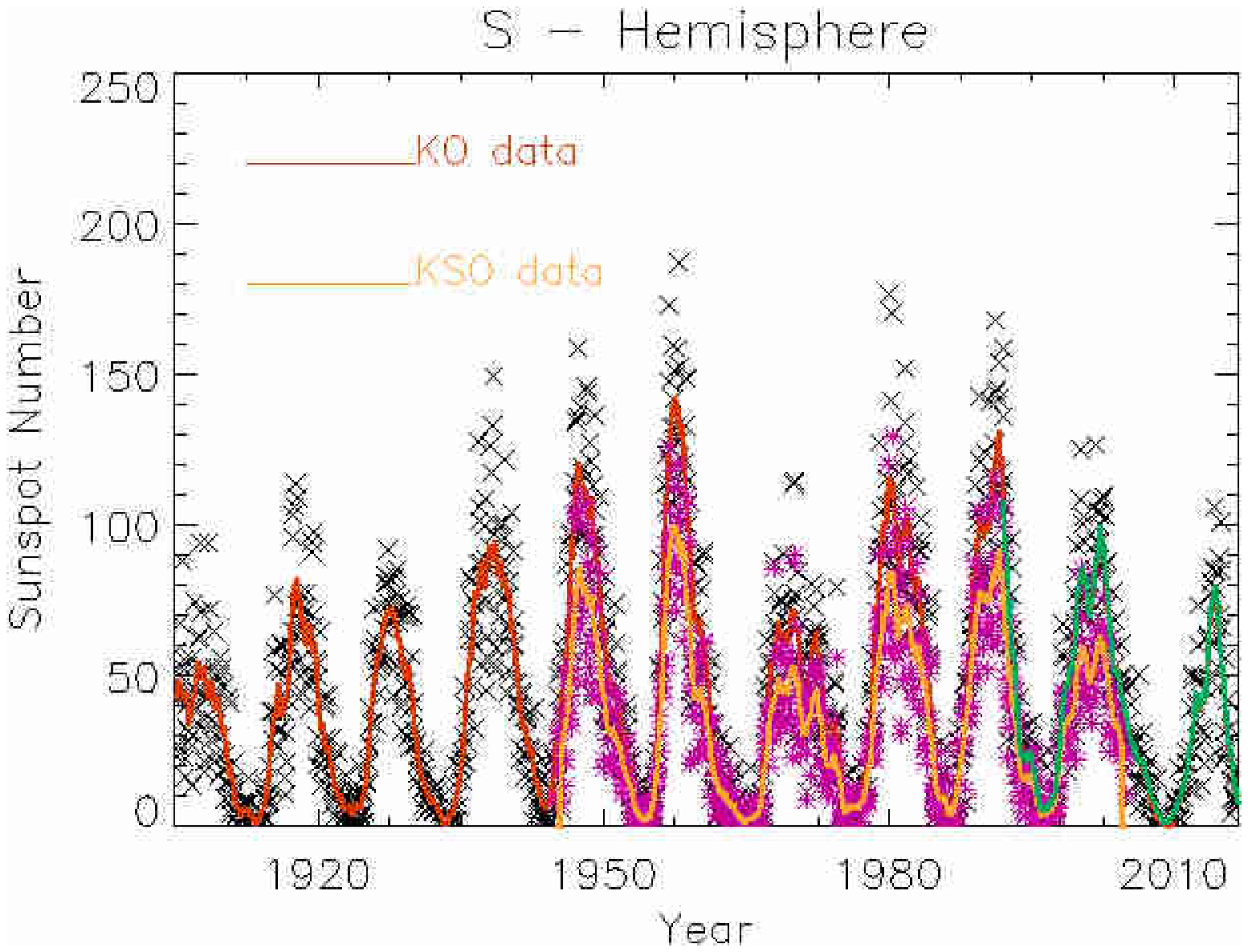} \\
\includegraphics[width=0.45\textwidth]{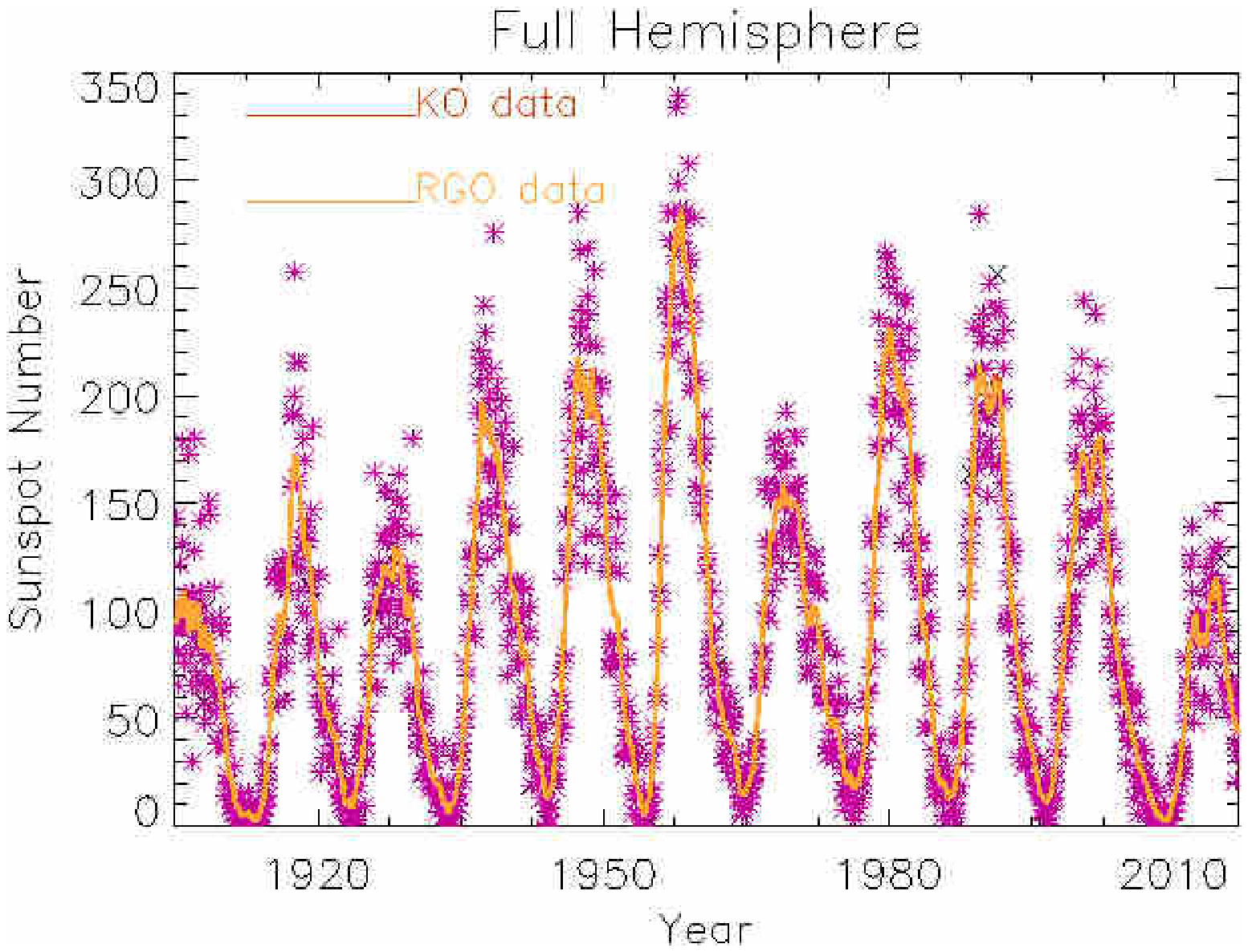} \\
\end{center}
\caption{Top: The monthly averaged and normalized sunspot number plotted for the Northern(left) and Southern (right) hemisphere staring from solar cycle 14 -- 24. The meaning of the symbols, colors and solid curves are the same as in Figure~\ref{fig:2}. Bottom: The monthly averaged and normalized total sunspot number. The data values are normalized to the RGO total sunspot number.}
\label{fig:5}
\end{figure*}

Figure~\ref{fig:5} shows the normalized sunspot number plot for the Northern (top-left), Southern (top-right) hemisphere and for the full-disk (bottom).
Clearly, the normalized KO sunspot number for the Northern and Southern hemisphere is larger than the KSO sunspot number but matches well with the SILSO 
sunspot number. It also matches well with the total sunspot number. The KO data set is very long covering about 11 cycles and hence it is very useful for further study on several scientific problems. 

\begin{figure}[!h]
\begin{center}
\includegraphics[width=0.45\textwidth]{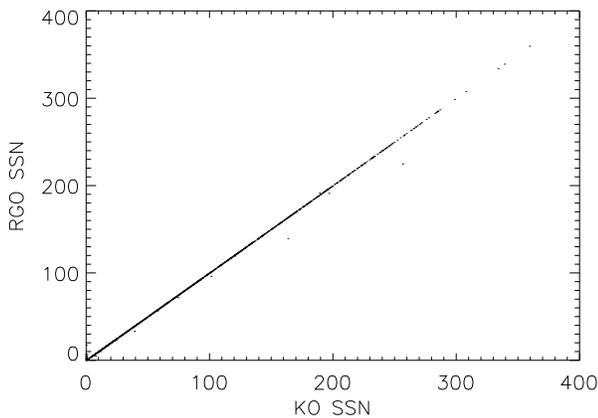} \\
\end{center}
\caption{A scatter plot between the Greenwich monthly averaged total sunspot number and the Kodaikanal monthly averaged and normalized total sunspot number. The obtained correlation coefficient is 0.99.}
\label{fig:6}
\end{figure}

Figure~\ref{fig:6} shows the scatter plot between the RGO and normalized KO sunspot numbers. Except for 8 points all of them lie on the straight line. This suggests that the long term data of KO sunspot number which is separately measured for the Northern and Southern hemisphere can be used in a similar way as RGO data in the future. For the present study, we show that even though the KO sunspot numbers are underestimated, its values well correlate with the RGO and hence able to calibrate it to the RGO data sets. 

\subsection{Sunspot Group Area}

\begin{figure}[!h]
\begin{center}
\includegraphics[width=0.45\textwidth]{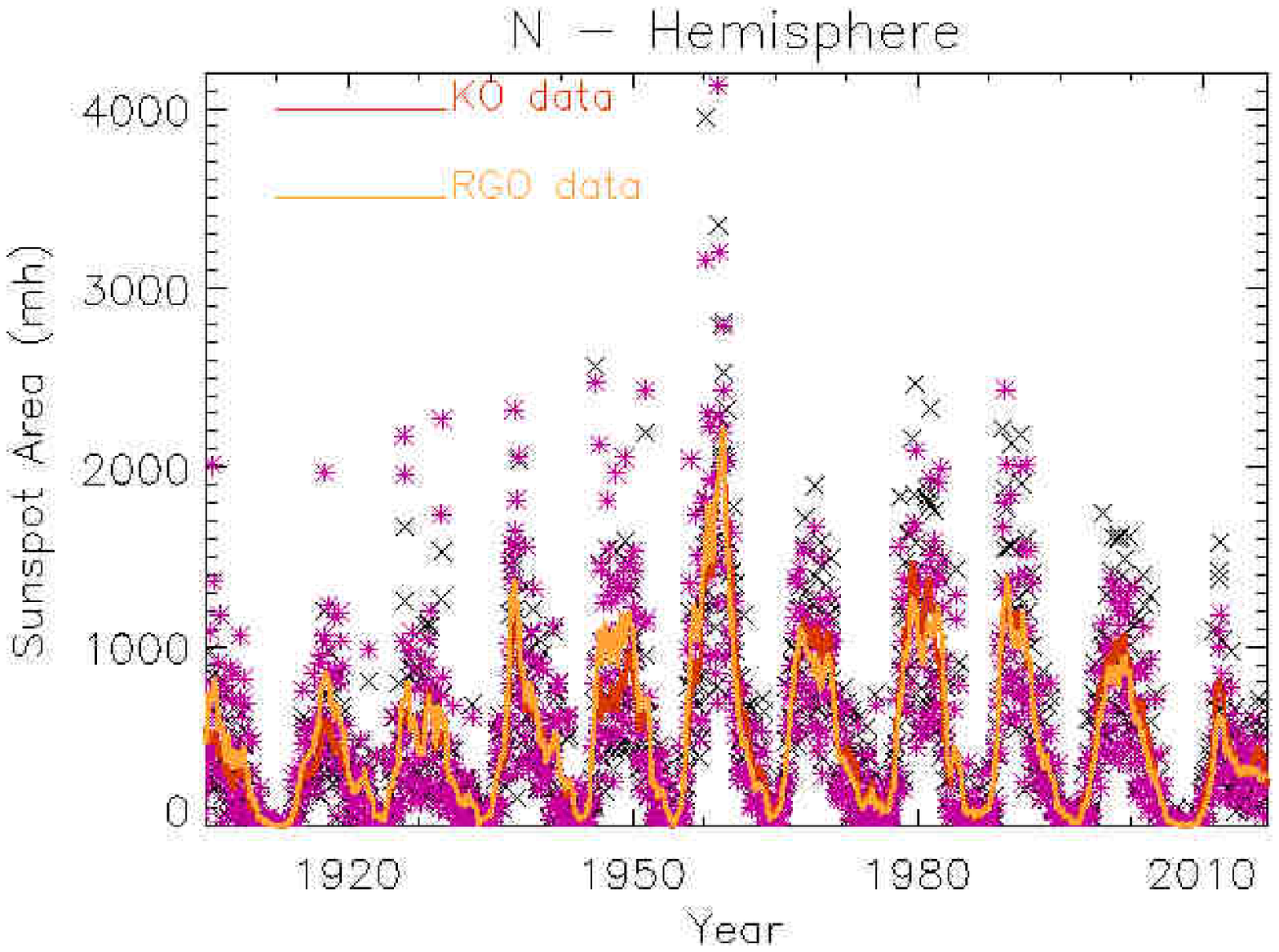} \\
\includegraphics[width=0.45\textwidth]{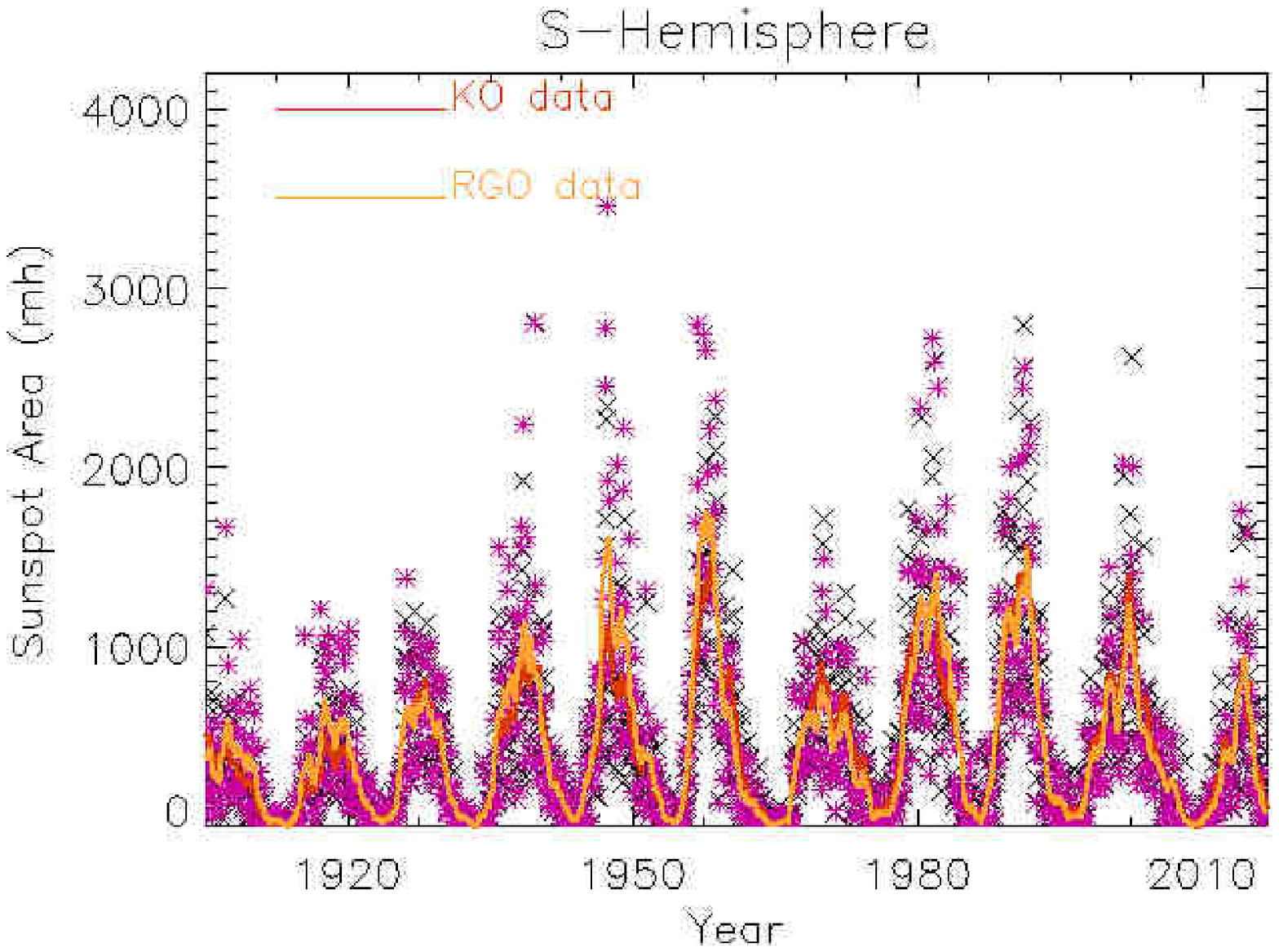} \\
\end{center}
\caption{The monthly averaged sunspot group area for cycle 14 through 24 is plotted for the Northern (top) and Southern (bottom) hemisphere respectively. The black crosses ($\times$) and pink star ($\ast$) symbols represent the Kodaikanal and Greenwich sunspot group areas respectively. The red and orange curve represents the 13-months smoothed KO and RGO sunspot group area respectively. The area is expressed in terms of a millionth of a hemisphere (mh).}
\label{fig:7}
\end{figure}

The sunspot group area is measured using the grid and the conversion table available at the Kodaikanal Observatory. The conversion table provides 
the values of the grid occupied by the sunspot group in terms of a millionth of a hemisphere. It also provides corrected values for the projection effect for the given heliocentric angle.  
The sunspot group area measured on the solar disk is given by,
\begin{equation}
A_{M} = \frac{2A_{s}10^{6}}{\pi D^{2} \cos(\rho)}
\end{equation}
where A$_{s}$ is the measured size of a sunspot group in the image. $\rho$ is the angular distance of the center of the sunspot group to the center of the solar disk. D is the diameter of the solar image. The area measured in a millionth of a hemisphere and it is corrected for the foreshortening effect. 

The sunspot group area observed in the Northern (top) and Southern (bottom) hemisphere is plotted in Figure~\ref{fig:7}. The black crosses represent the KO data and the pink star represents the RGO data. The red and orange curves represent the 13-months smoothed KO and RGO data.
The 13-months smoothed curve shows that the pattern of the solar cycle is similar in both the data set. The patterns of double or several peaks seen during the maximum period of the cycle appear to be similar in all the cycles. The amplitude of the cycles before cycle 19 is a little smaller and those appearing after cycle 19 is a little larger than the RGO's cycle amplitude. A similar trend is observed in the Southern hemisphere sunspot group area as well. This could be due to several factors such as change over from photographic plates to films and hence the change in the contrast of the sunspots in the image, a number of different observers who made the drawing, different observer who measured the sunspot area and change in the sketching pencil. All these will contribute to the error in the measurements systematically over time.

\begin{figure*}[!h]
\begin{center}
\includegraphics[width=0.5\textwidth]{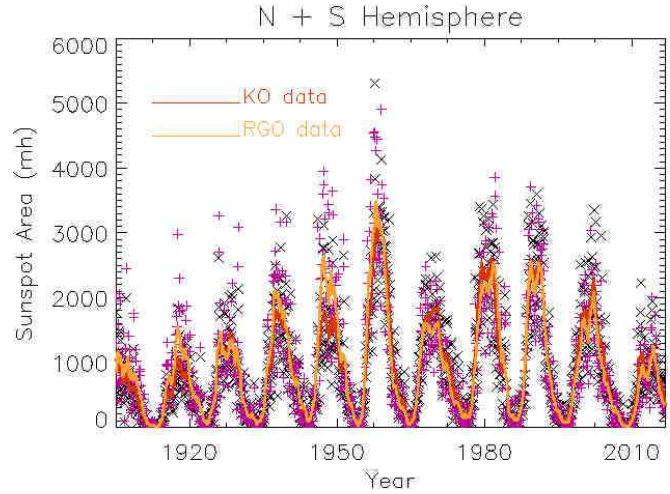} \\
\end{center}
\caption{The monthly averaged total sunspot group area is plotted against time. The symbols, colors and solid curves indicate respective information and it is the same as in other figures.}
\label{fig:8}
\end{figure*}

Figure~\ref{fig:8} shows the total sunspot area plot for cycle 14 to 24 measured over the whole hemisphere. Once again the black cross and pink colored star symbols represent the KO and RGO data sets respectively. The 13-months smoothed KO (red) and RGO (yellow) sunspot group area are also shown in the same plot. As in previous plots, the 13-months smoothed data the patterns of peaks occurred during the sunspot maximum are similar in both the data sets. The KO data shows systematically underestimated before cycle 19 and overestimated after cycle 19. However, the differences are small. In cycle 19 the peak occurred in the KO data at a different times than the RGO data. Because of this, the cycle shape has changed and it is shifted by a few months. This could be due to the error made by the observer in measuring the group area.  Even the 10 -- 15\% error in the measurement can shift the peak to some other time.  This has happened in the southern hemisphere sunspot group area measurements and is depicted in Figure~\ref{fig:7}(bottom). Similarly in cycle 18, the first peak is taller than the second in RGO, but in KO data all the peaks are at the same level. Likewise, the differences can be seen in cycle 16 and 21 as well.  

\begin{figure*}[!h]
\begin{center}
\includegraphics[width=0.5\textwidth]{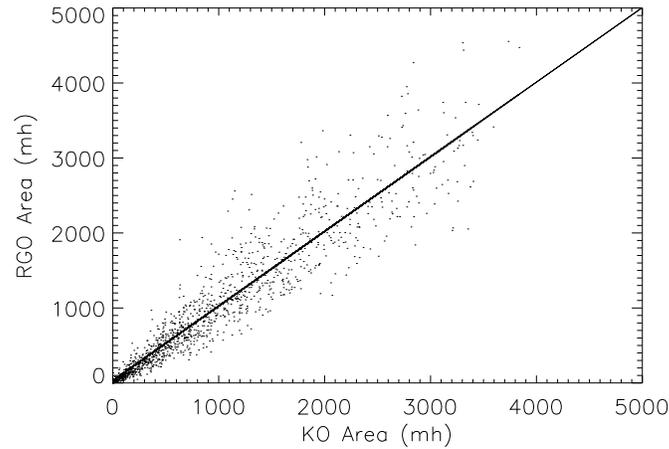} \\
\end{center}
\caption{The monthly averaged RGO sunspot group area is plotted against the KO sunspot group area. The solid line is a linear least-squares fit to the data points. The obtained correlation coefficient is 0.94.}
\label{fig:9}
\end{figure*}

In order to understand clearly how well the KO and RGO sunspot area are correlated, we plot a graph of RGO {\em versus} KO sunspot group area (Figure~\ref{fig:9}). It is found that they are correlated well with 0.94 as coefficient value. The scatter plot shows that the scatter is small at smaller group areas and it is a little large at larger sized sunspot groups. A gradient-expansion least-square fit to the data points gave the following relationship between the two data set as,

\begin{equation}
RGO\_A = 0.99(\pm 0.0095)KO\_A + 31.27(\pm 11.79) 
\end{equation}

The slope is very close to unity suggesting that both the sunspot group area measurements are very close to each other. Though the scatter is large at the larger area beyond 2000 millionth of a hemisphere, but the number of such points is small compared to the smaller sunspot groups. 

\subsection{Correlation Between Sunspot Number and Group Area}

\begin{figure*}[!h]
\begin{center}
\includegraphics[width=0.5\textwidth]{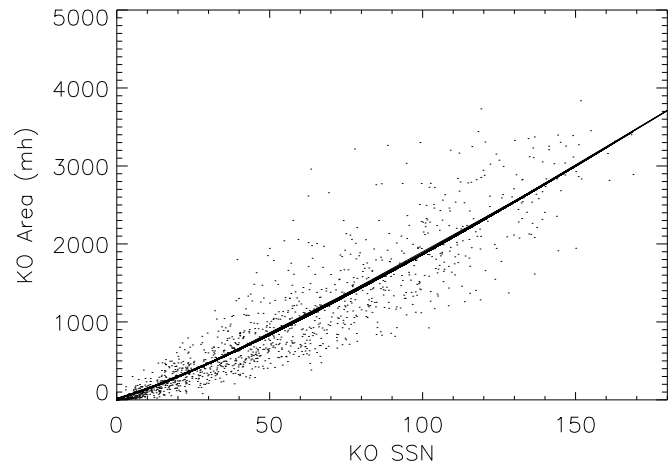} \\
\end{center}
\caption{A scatter plot between the monthly averaged sunspot area and sunspot number obtained from KO sunspot drawings. The solid curve is a gradient-expansion least-square fit to the data points. The obtained correlation coefficient is 0.91.}
\label{fig:10}
\end{figure*}

It is seen that in the RGO data, there is a very good correlation between the sunspot number and the area occupied by the sunspot group. We expect a similar trend in the KO data as well. To find that, we have plotted a graph between the KO sunspot group area and the sunspot number (Figure~\ref{fig:10}). We have used the monthly averaged data. The obtained correlation coefficient is 0.92. Once again the scatter is large for large sunspot group area. This could be happening during the sunspot maximum. From the plot, we found a relationship between the sunspot area and number as $A=8.15 \times SSN^{1.17} + 15.85$ which is
slightly different than quoted in \cite{1988assu.book.....Z}. 

\section{Summary and Discussions}
Kodaikanal Observatory has begun the observations of the sun about 115 years ago. The images are recorded in those times in the photographic plates/films and some of them continued until today. Using these different images obtained in different wavelengths the drawings of sunspots, plages, filaments, and prominences are made. Each of these features is identified and marked with different colors. A unique feature of KO drawings is that all the features are available in one image which is rare in other data sets. In this way, one can easily see the activity on the sun. These features are drawn on the 5$^{\circ}$ resolution Stonyhurst grid which makes it easy to identify where exactly the features located on the solar hemisphere. Otherwise just looking at the photographs it is not easy to identify the latitude and longitude of the features on the sun. 

All these features are drawn on the {\em sun chart} by either keeping the paper on the photographic plates or by projecting the image of the sun from the plates on to the paper. In this case, a precaution has been taken to match the grid circle with the image. The drawings are made as and when the photographic plates or films are developed. In this way, one can avoid the scratches and fungus which are present in the preserved photographic plates. 

In the past, several observatories around the world (\cite{2010LRSP....7....1H, 2015LRSP...12....4H} and references therein) have made the sunspot number measurements based on the photographic plates and also using drawings/{\em sun charts}.
They have followed Wolf's method to measure the relative sunspot numbers. In a similar way, we have also used the drawings made at KO to count the sunspot numbers. The magnitude of KO sunspot numbers is small compared to the RGO numbers at least by about 40\%. This can be easily corrected using the correction factor {\em k}. However, we have only one measurement by one or two observers. In this way using {\it k} value obtained from the scatter plot will not provide a complete solution. A better solution would be the one used by \cite{2002A&A...390..707T}.  The present KO sunspot number data is also normalized with respect to the RGO total sunspot number. This resulted in a very good correlation of the KO data with RGO including its cycle amplitude. The KO has covered 11-solar cycle and the normalized Northern and Southern sunspot number data will be very useful to study the North-South asymmetry in sunspot numbers and many more which has not been done for a long term data set simply because the data is not available (see also; \cite{2002A&A...390..707T, 2006A&A...447..735T, 2019SoPh..294...79L}). 

The manual measurement of the sunspot group area is time-consuming, but we would like to compare our results with the one done in the past with RGO data whose measurement is made in a similar method. Comparing our measurements with the RGO sunspot area measurements we find that for about 5 cycles the area is under-estimated by about 10\% and for the later 5-solar cycles it is overestimated by about 10\%. This could be because of two different observers whose measurements will not match exactly. This could be also due to the exchange of photographic plates and films thereby contrast of the recorded feature changes. Before 1960, at KO the observations were recorded with photographic plates and later with films. Definitely, there is a change in the contrast of the image obtained in plates and films. The image quality is better in plates. With the better contrast images, but with low spatial resolution the area was underestimated and later with the low contrast images the area was overestimated. Even the seeing could have deteriorated the images in the later case and hence the overestimation of the area by about 10\%. 
 
There are many ways of identifying the sunspots in a group, counting the number of sunspots in a group and on the visible disk of the sun, and the area occupied by the sunspot as well as groups. This information can be extracted from the photographic plates or from sunspot drawings. There are several software tools developed to extract this information from KO photographic plates \citep{2017ApJ...835..158M, 2017A&A...601A.106M}. A semi-manual method was also used to extract the information such as tilt angle \citep{1999SoPh..189...69S}, rotation rate \citep{1999SoPh..186...25H}, etc. In this paper, we presented our results extracted using a completely manual approach which is similar to the one used at the RGO. These {\em sun charts} are now being digitized and in future, one can extract more information from these using automated methods. This may provide another additional data set for comparison.

\section*{Acknowledgment}
We thank the referee for fruitful comments.
The data used here is proprietary of the Kodaikanal Solar Observatory. We thank all the observers who made the observations over the period of 115-years and also made the drawings and sketches of all the features on the sun chart.  Special thanks to Anitha, Jayalakshmi, Avanthika, Madhumitha for cross examing the sunspot numbers. SILSO sunspot number data used in this study are provided by the Royal Observatory of Belgium, Brussels.

\bibliographystyle{spr-mp-nameyear-cnd.bst}
\bibliography{reference.bib}

\end{document}